\documentclass[aps,prb,twocolumn,superscriptaddress,showpacs,preprintnumbers,amsmath,amssymb,floatfix]{revtex4}
\usepackage{graphicx}
\usepackage{dcolumn}
\usepackage{bm}
\graphicspath{{./figs/}}

\begin{document}

\title{All-strain based valley filter in graphene nanoribbons using snake states}

\author{L. S. Cavalcante}\email{lucaskvalcante@fisica.ufc.br}
\affiliation{Universidade Federal do Cear\'a, Departamento de
F\'{\i}sica Caixa Postal 6030, 60455-760 Fortaleza, Cear\'a, Brazil}
\author{A. Chaves}\email{andrey@fisica.ufc.br}
\affiliation{Universidade Federal do Cear\'a, Departamento de
F\'{\i}sica Caixa Postal 6030, 60455-760 Fortaleza, Cear\'a, Brazil}
\affiliation{Department of Chemistry, Columbia University, 3000 Broadway, 10027 New York, NY}
\author{D. R. da Costa}\email{diego_rabelo@fisica.ufc.br}
\affiliation{Universidade Federal do Cear\'a, Departamento de
F\'{\i}sica Caixa Postal 6030, 60455-760 Fortaleza, Cear\'a, Brazil}
\affiliation{Department of Physics, University of Antwerp, Groenenborgerlaan 171, B-2020 Antwerp, Belgium}
\author{G. A. Farias}\email{gil@fisica.ufc.br}
\affiliation{Universidade Federal do Cear\'a, Departamento de
F\'{\i}sica Caixa Postal 6030, 60455-760 Fortaleza, Cear\'a, Brazil}
\author{F. M. Peeters}\email{francois.peeters@uantwerpen.be}
\affiliation{Department of Physics, University of Antwerp, Groenenborgerlaan 171, B-2020 Antwerp, Belgium}
\affiliation{Universidade Federal do Cear\'a, Departamento de
F\'{\i}sica Caixa Postal 6030, 60455-760 Fortaleza, Cear\'a, Brazil}

\begin{abstract}
A pseudo-magnetic field kink can be realized along a graphene nanoribbon using strain engineering. Electron transport along this kink is governed by snake states that are characterized by a single propagation direction. Those pseudo-magnetic fields point towards opposite directions in the $K$ and $K'$ valleys, leading to valley polarized snake states. In a graphene nanoribbon with armchair edges this effect results in a valley filter that is based only on strain engineering. We discuss how to maximize this valley filtering by adjusting the parameters that define the stress distribution along the graphene ribbon. 
\end{abstract}

\maketitle

\section{Introduction}

The advent of graphene \cite{Geim, CastroNetoReview} not only represented the beginning of a new era of atomically thin materials, with potential technological applications in future electronic and photonic devices, but also brought the possibility of observing several novel phenomena due to its unique band structure, consisting of Dirac cones in points labeled as $K$ and $K'$ in its first Brillouin zone. In fact, the existence of two inequivalent cones is of special importance, since it enables a new degree of freedom to be explored in novel valley-tronic devices. 

Several suggestions have been made to harvest valley polarization in graphene: Rycerz et al. \cite{ValleyFilter1} demonstrated that specific combinations of armchair and zigzag edges in a monolayer graphene ribbon lead to efficient valley filtering. Non-uniform substrate induced masses can also be used to obtain valley polarization, as shown in Refs. [\onlinecite{Zarenia, Massoud}]. As for bilayer graphene, valley filtering can be obtained by specific configurations of external potentials, \cite{Diego} or boundaries with monolayer graphene regions. \cite{Koshino, Pratley} On the other hand, recent studies have demonstrated that pseudo-magnetic fields can be induced in graphene by specific strain configurations and, since these fields point towards opposite directions in different Dirac cones \cite{Guinea}, several suggestions of strain-based valley filters have been proposed in the literature. Most of these proposals involve combinations of the strain induced fields with applied magnetic and electric fields.\cite{TonyLow, SGTan, Zhai, Myoung, Zhao} Indeed, strain-based valley filters are specially interesting, because of graphene's ability to withstand large mechanical stress. \cite{Strain} Very large pseudo-magnetic fields have been experimentally observed in e.g. naturally formed bubbles in a monolayer graphene system on a Pt substrate. \cite{CastroNeto300T}

\begin{figure}[!b]
\centerline{\includegraphics[width=\linewidth]{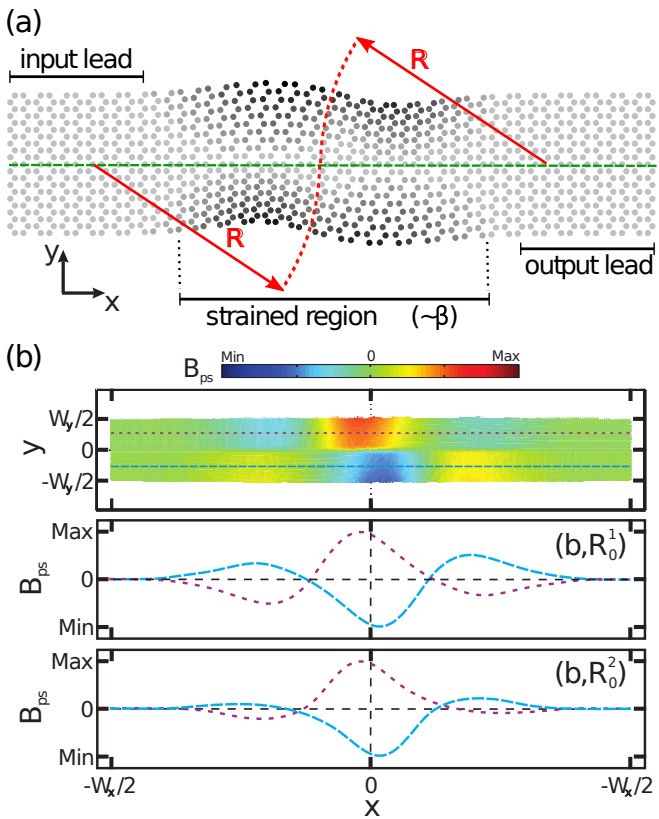}}
\caption{(Color online) (a) Sketch of the strained graphene ribbon. Strain is defined by two circles of radius $R$ (red dashed), which smoothly decay to zero towards the input and output leads, and the width of the strained region is defined by $\beta$. The color map indicates smaller (brighter regions) and larger (darker regions) local displacements. Central atoms (green dashed line along the $x$-axis) are always unstrained in this configuration. (b) (top) Contour plot of the induced pseudo-magnetic field profile for a representative strain configuration characterized by the parameters ($\beta$, $R_0$). (bottom) A cross-view of the pseudo-magnetic field along the lines placed in $y > 0$ (purple short-dashed) and $y < 0$ (blue long-dashed) regions of the system in the top panel, for two different maximum radii $R^2_0 > R^1_0$.} 
\label{fig:sketch}
\end{figure}

In this paper, we propose a very different valley filter device made of a single graphene layer \cite{Diego1} that does not depend on specific edge configurations, \cite{ValleyFilter1} substrate induced masses,\cite{Zarenia} or external magnetic fields \cite{TonyLow, Chaves}. Yet, it is all based on a particular kind of strain induced in a monolayer graphene nanoribbon, that provides a pseudo-magnetic kink barrier along the ribbon width. Such strain is expected to be attainable by using flexible substrates in combination with e.g. appropriate piezos or contacts that deform when cooled down. \cite{TonyLow,Strain,Downs, Korhonen} The valley polarization originates from a combination of (i) the uni-directional motion of snake states along the pseudo-magnetic kink, and (ii) the opposite direction of the pseudo-magnetic field felt in the $K$ and $K'$ valleys. This combination leads to electrons moving in single orbits propagating in opposite directions in the two different valleys. In order to verify the efficiency of such valley filtering device, we calculate the transmission probability of wavepackets through this structure within the tight-binding model. Our results demonstrate that a valley polarization efficiency up to $90\%$ can be reached, provided specific conditions are met by the system, as we will discuss in what follows.

\section{Theoretical model for the strain}

Our system consists of a monolayer graphene nanoribbon with width $W_y \approx 6387$ \AA\ and  length $W_x \approx 2214$ \AA\, corresponding to $1801 \times 3000$ carbon atoms, that is mechanically strained in a specific configuration, as sketched in Fig. \ref{fig:sketch}: along a certain region of length $\beta$, the ribbon is distorted into two circular arcs of radius $R$, in opposite directions. Such in-plane circular bending is obtained by defining the displacement of the atomic sites as\cite{Strain}
\begin{equation}
 u_x(x,y) = (R+x)\cos \left[\frac{2y}{W_y}\arcsin\left(\frac{W_y}{2R}\right)\right]-y,
\end{equation}
and
\begin{equation}
 u_y(x,y) = (R+x)\sin \left[\frac{2y}{W_y}\arcsin\left(\frac{W_y}{2R}\right)\right]-R-x,
\end{equation}
where $u_{x,y}$ is the in-plane lattice distortion due to strain and the radius has its sign reversed at the $y = 0$ axis, i.e. $R = |R|\left(2\theta(y)-1\right)$, with $\theta(y)$ being the step function. A sharp transition between strained and unstrained regions of the ribbon would be clearly impossible, since it would lead to unrealistically large atomic distances in the vicinity of the transition region, specially for small $R$. Therefore, we consider a smooth (Gaussian) variation of the curvature $K_R = 1/R = (1/R_0)e^{(-j^2/\beta^2)}$, where $j$ is the index of the column to which a given atomic site belongs in the lattice, \cite{Andrey} the length is described by the dimensionless parameter $\beta$, while $R_0$ provides the maximum radius of the curve (namely, at the central column of atoms, where $x_{i,j} = 0$ in the absence of strain).

In general, strain effects on the electronic properties of graphene can be mapped into the analogous problem of an electron under a pseudo-magnetic field distribution,\cite{Amorim} whose magnitude and orientation may oscillate over the space, thus making the production of a local non-zero pseudo-flux challenging. \cite{Wehling} However, it has been recently demonstrated \cite{Guinea, Strain} that an in-plane circular distortion, as the one proposed here, deforms the Brillouin zone, shifting the Dirac cones with respect to each other, just like when an uniform magnetic field is applied perpendicular to the graphene plane, leading to $\bf{K}$$\rightarrow$$\bf{K}$$+2\pi$$\bf{A}$$/\Phi_0$, where $\Phi_0 = e/h$ is the flux quantum. Such lattice distortion changes the hopping energies and thus induce an effective vector potential \cite{Dean1, Dean2}
\begin{equation}
A_x + iA_y = \frac{1}{e v_F} \sum_{\bf{\delta a}_{ij}} \delta \tau_{ij} \mbox{~} e^{-i \bf{K}\cdot \bf{\delta a}_{ij}} 
\end{equation}
where $\bf{\delta a}_{ij}$ is the vector distance between the adjacent atoms $i$ and $j$ in the strained lattice, $v_F$ is the Fermi velocity, $\delta \tau_{ij}$ is the difference between the strained and unstrained hopping energies, and the pseudo-magnetic field is given by $\bf{B_{ps}} = \nabla \times \bf{A}$. Moreover, the distortion in different directions for $y > 0$ and $y < 0$ provides a pseudo-magnetic kink barrier, with pseudo-magnetic field regions that change sign at $y = 0$. A schematic example of the pseudo-magnetic field distribution induced by such strain configuration is presented in Fig. \ref{fig:sketch}(b) for a representative set of parameters ($\beta$, $R_0$). We point out that the sample considered in Fig. \ref{fig:sketch}(b) is much smaller than the one investigated throughout this paper, since calculating and plotting a vector potential distribution along the 1801 $\times$ 3000 atomic sites of our actual sample requires high computational costs. Therefore, the pseudo-magnetic field in Fig. \ref{fig:sketch}(b) is discussed here only in a qualitative way. The pseudo-magnetic field is found to be zero at input and output leads, where the lattice displacements vanish, and assume its minimum and maximum values along the ribbon width around $x = 0$, where the strain is maximum. Two additional kinks are also consistently observed on the left and right sides of this main central kink. They are however much smaller than the central one and, thus, do not play an important role in the valley filtering process, as we will demonstrate further on. In fact, we observe that as we increase $R_0$, these additional kinks become even lower as compared to the main kink, so that their importance for the transport properties of the actual sample studied throughout the paper (with larger $R_0$) is negligible. This can be verified by comparing the bottom panels in Fig. \ref{fig:sketch}(b) for two different maximum radii $R^2_0 > R^1_0$ assuming a fixed width for the strained region $\beta$.

\section{Snake states along a magnetic field kink}

Keeping with the analogy between this strain configuration and a magnetic field kink, let us first calculate the energy dispersion along the ribbon in the presence of such a magnetic barrier. We assume an inhomogeneous magnetic field $\bf{B}$ $= B \hat{z}$ that depends only on the transversal coordinate, given by $B(y) = B(\theta(y) - \theta(-y))$. Low energy electrons in graphene exhibit a linear energy dispersion, so that they behave as massless Dirac-Weyl fermions, thus, obeying the Dirac equation:
\begin{equation}
 {\bf{\sigma}}\cdot\left(-i{\bf{\nabla}} + \frac{e}{\hbar}\bf{A}\right)\Psi = \bar E\Psi,
 \label{Dirac}
\end{equation}
with energy $E = \hbar v_{F}\bar E$ around the $K$ valley (a similar analysis \cite{CastroNetoReview} can be made for electrons in $K'$). 

Defining the vector potential in the Landau gauge, $\textbf{{A}} = \bar{A}(y)\hat{e}_{x}$, with $\bar A(y) = -B(y)y$, the general solution for the wavefunction with translational invariance in the $x$-direction is $\Psi(x,y) = \psi(y)e^{ikx}$. Therefore, we obtain from Eq. (\ref{Dirac})
\begin{eqnarray}
\begin{split}
& \left( \begin{array}{cc}
0 & k - \partial_y + A(y) \\ 
k + \partial_y + A(y) & 0
\end{array} \right) 
\left( \begin{array}{cc}
\psi_1(y) \\ 
\psi_2(y)
\end{array} \right) \\ 
&= \bar E \left( \begin{array}{cc}
\psi_1(y) \\ 
\psi_2(y)
\end{array} \right),
\end{split}
\end{eqnarray}
where $A = \frac{e}{\hbar}\bar A$. This leads us to decoupled equations for each component: for instance, for the upper spinor component,
\begin{equation}
 \left\{\partial^2_y + \frac{e}{\hbar}B(y) - \left[k - \frac{e}{\hbar}B(y)y\right]^2 + \bar E^2 \right\} \psi_1(y) = 0.
\end{equation}
If we use the magnetic length $l_b = \sqrt{\hbar / e|B|}$ as the unit of distance, we obtain
\begin{equation}
 \left\{\partial^2_y + sgn(B(y)) - \left[kl_b - sgn(B(y))y\right]^2 + \epsilon^2 \right\} \psi_1(y) = 0.
\end{equation}
where $\epsilon = \bar E l_b = E l_b \big/ \hbar v_F$.

The energy dispersion along the $y$-direction is obtained quasi-analytically by solving this equation in terms of parabolic cylinder functions \cite{Gosh}, $D_{p}(q)$. Notice the magnetic field $B(y)$ is piecewise constant, hence, one can separate solutions for each region as
\begin{equation}
\psi_{B>0}(y) = \sum_\pm a_\pm \left( \begin{array}{cc}
D_p(\pm q) \\ 
\mp\frac{\sqrt{2}}{i\epsilon}D_{p+1}(\pm q)
\end{array} \right),
\label{mais}
\end{equation}
\begin{equation}
 \psi_{B<0}(y) = \sum_\pm a_\pm \left( \begin{array}{cc}
D_{p+1}(\pm q) \\ 
\pm\frac{\sqrt{2}}{i\epsilon}(p+1)D_p(\pm q)
\end{array} \right),
\label{menos}
\end{equation}
where $q = \sqrt{2}\left[sgn(B)k l_b\right]$, and $p = \frac{\epsilon^2}{2} - 1$. 

The continuity of the wavefunction and its derivatives at these regions provides boundary conditions that lead to quantization of the energy of the system. Eqs. (\ref{mais}) and (\ref{menos}) represent solutions for the first region of the system, but they also can express solutions for the second region by replacing the coefficients $a_{\pm} \rightarrow c_{\pm}$.

In order that the wavefunctions are normalizable we demand $a_+ = c_{-} = 0$. Then, the boundary condition at $y = 0$ gives the equation that generates the energy quantization condition. Therefore, we obtain
\begin{equation}
 \left(\frac{i\epsilon}{\sqrt{2}} v^2 -  \frac{\sqrt{2}}{i\epsilon} (p+1) u^2\right) = 0,
\end{equation}
where the functions are given by $u = D_p (-\sqrt{2} k l_b)$, and $v = \frac{\sqrt{2}}{i\epsilon} D_{p+1}(-\sqrt{2} k l_b)$.
Notice that these results for the wavefunction are very closely related to the one for a magnetic kink profile in a normal $2D$ semiconductor. \cite{Reijniers} 

\begin{figure}[!t]
\centerline{\includegraphics[width=0.975\linewidth]{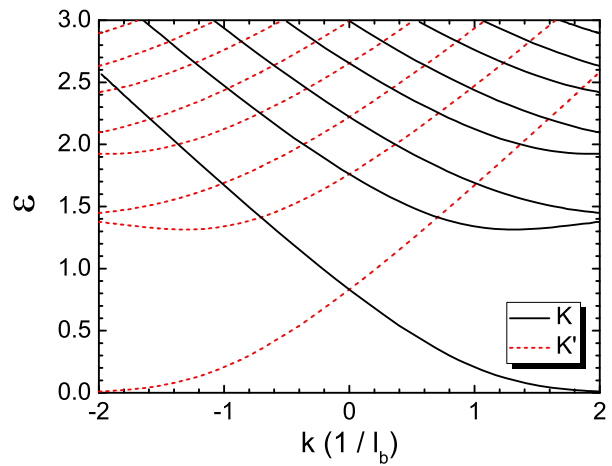}}
\caption{(Color online) Energy states for a Dirac particle in the presence of a (pseudo-)magnetic kink. Results for $K$ and $K'$ valleys are the same for an external magnetic field, whereas for a pseudo-magnetic field, the $K'$ spectrum (red dashed curves) differs from that from the $K$ valley (black solid curves).} \label{fig:spectrum}
\end{figure}

Numerical results for this system are illustrated in Fig. \ref{fig:spectrum}, where we observe an asymmetry in the energy bands along the kink with respect to the $k_x = 0$ axis. Namely, the energy states (predominantly) monotonically decrease with $k_x$, implying a negative velocity $v = (1/\hbar)dE/dk$, that eventually converges to zero as $k = k_x \rightarrow +\infty$. It is straightforward to verify that by inverting the sign of the magnetic field kink, this figure is reflected with respect to the $k_x = 0$ axis, and the propagation velocities are then predominantly positive. The physical interpretation of this result has its basis on the existence of snake states\cite{Oroszlany, Mele} that propagate along the kink, which can also be inferred from a simple classical analysis of this problem, involving Lorentz force, cyclotron orbits and the right-hand rule. Nevertheless, this result is of special importance in the context of pseudo-magnetic kinks discussed here: since the strain-induced pseudo-magnetic field points towards opposite directions in the different Dirac cones $K$ and $K'$, electrons in each cone will propagate in opposite directions in the pseudo-magnetic kink proposed here, thus yielding an efficient valley filtering process, as we will verify with our numerical results afterwards.

\section{Wavepacket propagation method}

In order to investigate the transport properties in our strained graphene, we will use a wavepacket propagation method. A comparison between this kind of method and those based on Green's function formalism can be found in Ref. [\onlinecite{Kramer}]. The advantage of using this approach is in the possibility of observing the trajectories of the wave packet describing the electron propagating across the scattering (strained) region, which reveals important information about the physics behind any unusual behavior of the current through the system, being due to e.g. inter-valley scattering, skipping orbits, snake states, etc., as we will demonstrate in the following Section.

We use a Hamiltonian within the tight-binding model
\begin{equation}
 H_{TB} = - \sum_{i,j} \tau_{ij}c^\dagger_i c_j + h.c.,
\end{equation}
where the operator $c^\dagger_i (c_i)$ creates (annihilates) an electron on site $i$, and $\tau_{ij}$ is the hopping energy between adjacent atoms $i$ and $j$ (nearest-neighbors), that depends on the distance $\delta a_{ij}$ between them according to \cite{Guinea}
\begin{equation}
 \tau_{ij} \rightarrow \tau_{ij} \left( 1 + \frac{2\delta a_{ij}}{a_0}\right).
\end{equation}

We consider an initial Gaussian wavepacket
\begin{equation}\label{eq.wave}
 \Psi(x,y) = N\exp\left[-\frac{(x-x_0)^2 + (y-y_0)^2}{2\sigma^2} + ik_x x + ik_y y\right],
\end{equation}
where $N$ is a normalization factor and calculate its time evolution using the split-operator method, \cite{Chaves, Andrey} in which the time-evolution operator for the Hamiltonian $H = H_i + H_j$ is split as
\begin{equation}
 \Psi^{t+\Delta t}_{ij} = e^{-\frac{i}{2\hbar} H_i \Delta t} e^{-\frac{i}{\hbar} H_j \Delta t} e^{-\frac{i}{2\hbar} H_i \Delta t} \Psi^t_{ij},
\end{equation}
where $H_{i(j)}$ is the term of the tight-binding Hamiltonian $H$ that corresponds to a horizontal (vertical) hopping between atomic sites
\begin{eqnarray}
H_i |i,j\rangle = \tau^{\prime}_{ij} |i,j+1\rangle + \tau^{\prime\prime}_{ij}|i,j-1\rangle \\
H_j |i,j\rangle = \tau_{ij} |i+1,j\rangle + \tau_{ij}|i-1,j\rangle.
\end{eqnarray}
Notice that for the horizontal term $H_i$, one has to differentiate between hoppings to the right and left neighbouring sites, since in the honeycomb lattice, each site has only horizontal hops to one side. Hence, $\tau^{\prime}_{ij} = \tau_{ij} \implies \tau^{\prime\prime}_{ij} = 0$ and $\tau^{\prime\prime}_{ij} = \tau_{ij} \implies \tau^{\prime}_{ij} = 0$.

The advantage of such splitting lies in the fact that these operators can be represented by tridiagonal matrices, that are easily handled by standard computational routines. The wavefunction after a single time step $t + \Delta t$ is then obtained in three steps
\begin{equation}
 \eta_{ij} = e^{-\frac{i}{2\hbar} H_i \Delta t}\Psi^t_{ij},
\end{equation}
\begin{equation}
 \xi_{ij} = e^{-\frac{i}{2\hbar} H_j \Delta t} \eta_{ij},
\end{equation}
\begin{equation}
 \Psi^{t+\Delta t}_{ij} = e^{-\frac{i}{2\hbar} H_i \Delta t}\xi_{ij}.
\end{equation}
Each of these equations is re-written using the Cayley form for the exponentials, e.g.
\begin{equation}
 \left(1+\frac{i\Delta t}{4\hbar} H_i \right) \eta_{ij} \approx \left(1-\frac{i\Delta t}{4\hbar} H_i \right) \Psi^t_{ij},
\end{equation} 
and the remaining tridiagonal matrix equation is then numerically solved by standard computational routines. \cite{NumericalRecipes}

For our study, we used a wavepacket width of $\sigma = 300$ \AA\ and its wave vector $\vec k$ has a modulus of $k = 0.06$ \AA$^{-1}$, unless otherwise explicitly stated in the text. Using the linear spectrum approximation for low-energy electrons in graphene, in which $E = \hbar v_f k$, the wavepacket energy is estimated to be $E = 343$ meV. Besides, as we intend to demonstrate the valley polarization of the wavepacket, we place it in different valleys in reciprocal space by shifting the wave vector towards the two inequivalent Dirac points: 
\begin{equation}
 k_x \leftarrow |k|,\ \
 k_y \leftarrow \pm \frac{4\pi}{3\sqrt{3}a},
\end{equation}
where the positive (negative) sign refers to a displacement towards the $K$ ($K'$) point of the Brillouin zone, and $a \approx 1.42$ \AA\ is the inter-atomic distance.

As the Gaussian wavepacket propagates, we calculate the probability of finding the electron before ($P_1$), within ($P_2$), and after ($P_3$) the strained region, as the integral of the square modulus of the wavepacket, taken within the intervals $-3,000$ \AA\ $\leq x \leq$ $-400$ \AA, $-400$ \AA\ $\leq x \leq$ $400$ \AA, and $400$ \AA\ $\leq x \leq$ $3,000$ \AA, respectively. Transmission probabilities are assumed to be the converged value of $P_3$ as $t \rightarrow \infty$. Besides, we keep track of the wavepacket trajectories by calculating the average value of the position, ($\langle x \rangle$, $\langle y \rangle$), at each time step.

The armchair edges of the ribbon do not support edge states, therefore, modelling the electron propagating through the system as a wavepacket, whose tails do not reach the ribbon edges, is justified. Moreover, any improvement to come from other calculation methods, involving e.g. plane waves, scattering matrices and the Landauer-Buttiker formalism, would lead to rather quantitative corrections to our results, while the qualitative behavior of the system and the proof-of-concept of valley filtering with a pseudo-magnetic kink proposed here, which are the main goals of this work, would still hold, since they are based on more fundamental physical properties of the proposed structure, as we will discuss in what follows.

\section{Results and discussion}

\begin{figure}[!t]
\centerline{\includegraphics[width=0.95\linewidth]{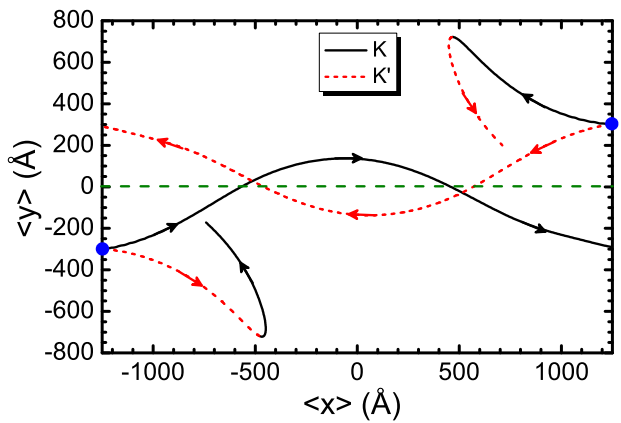}}
\caption{(Color online) Examples of calculated trajectories of electron wavepackets propagating with momenta around $K$ (black solid curves) and $K'$ (red dashed curves) valleys, starting at ($x,y$) points (indicated by blue solid dots) given by ($1250$ \AA, $300$ \AA) and ($-1250$ \AA,$-300$ \AA), respectively. Arrows indicate the direction of propagation along the trajectories.} \label{fig:snakes}
\end{figure}

\begin{figure}[!b]
\centerline{\includegraphics[width=\linewidth]{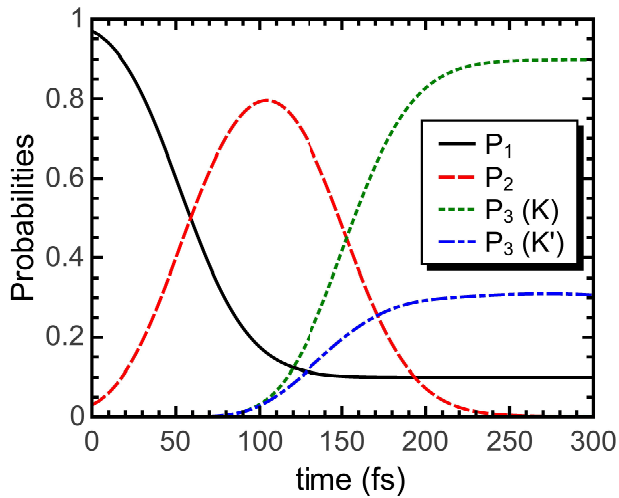}}
\caption{(Color online) Probability densities, as a function of time, of finding the electron before ($P_1$), within ($P_2$), and after ($P_3$) the $\beta = 900$ and $R_0 = 10,000$ \AA\ strained region, for a wavepacket with $k = 0.06$ \AA$^{-1}$ around the $K$ point of the Brillouin zone. Results for $P_3$ considering a wavepacket around $K'$ are shown for comparison.} \label{fig:prob}
\end{figure}

The existence of snake states in such a strained graphene lattice, as due to the induced pseudo-magnetic fields, is confirmed by the trajectories drawn in Fig. \ref{fig:snakes} of the center-of-mass of a $k = 0.06$ \AA$^{-1}$ wavepacket propagated in time through the system described by Fig. \ref{fig:sketch}(a), assuming $\beta \rightarrow \infty$ and $R_0 = 10^{4}$ \AA, as a test case. If this wavepacket has momentum around the Dirac cone $K$ (black solid curves) and propagates from left to right, starting at the bottom-half of the ribbon, its trajectory is deflected by the pseudo-magnetic Lorentz force towards the top-half, where it is deflected downwards again by the opposite pseudo-magnetic field, thus performing a snake-like trajectory. If this same packet has momentum around the $K'$ cone (red dashed curves), it is deflected downwards and eventually repelled from the strained region. If this packet starts from the top-half instead, both curves are just mirror-reflected with respect to the $\langle y \rangle$ = 0 axis of Fig. \ref{fig:snakes}, and the situation remains the same. Conversely, if the wavepacket propagates from right to left, it is the $K'$ packet that draws a snake trajectory, whereas the $K$ packet is reflected. One could think that wavepackets deflected towards the edges of the system (i.e., further away from its center) would be reflected by the ribbon edges, perform skipping orbits, and eventually pass through the strained region. However, since the ribbon has armchair edges, reflected wavepackets are scattered to the other Dirac cone, where the pseudo-magnetic field is opposite, thus the skipping orbit follows the opposite direction and the wavepacket comes back anyway. \cite{Diego1} This non-propagating edge state is emphasized in Fig. \ref{fig:snakes} for a wavepacket that started at the bottom-half (top-half) of the ribbon and around $K'$ ($K$) Dirac valley. In this way, one completely avoids the problem of having valley mixing of the snake and edge states at the end of the ribbon, which would otherwise occur in the case of propagating edge states e.g. in the presence of an external applied magnetic field. Valley mixing by scattering at the contacts can also be further suppressed by using graphene electrodes. \cite{Jo} 

Notice, however, that the present proposal will not work very efficiently for zigzag graphene nanoribbons, where such inter-valley edge scattering does not occur and skipping orbits are allowed to propagate at the zigzag edges. Although this represents a limitation of the proposed system, fabrication techniques have been advancing fast, and armchair graphene nanoribbons with very high edge quality have already been experimentally demonstrated. \cite{Kimouche} 

Results in Fig. \ref{fig:snakes}, thus, allow us to conclude that electrons in $K$ ($K'$) cones in such a strained \textit{armchair} graphene ribbon can only propagate towards the right (left). Analogously, if the strain configuration is inverted, trajectories drawn by $K$ and $K'$ packets are switched.

\begin{figure}[!t]
\centerline{\includegraphics[width=0.9\linewidth]{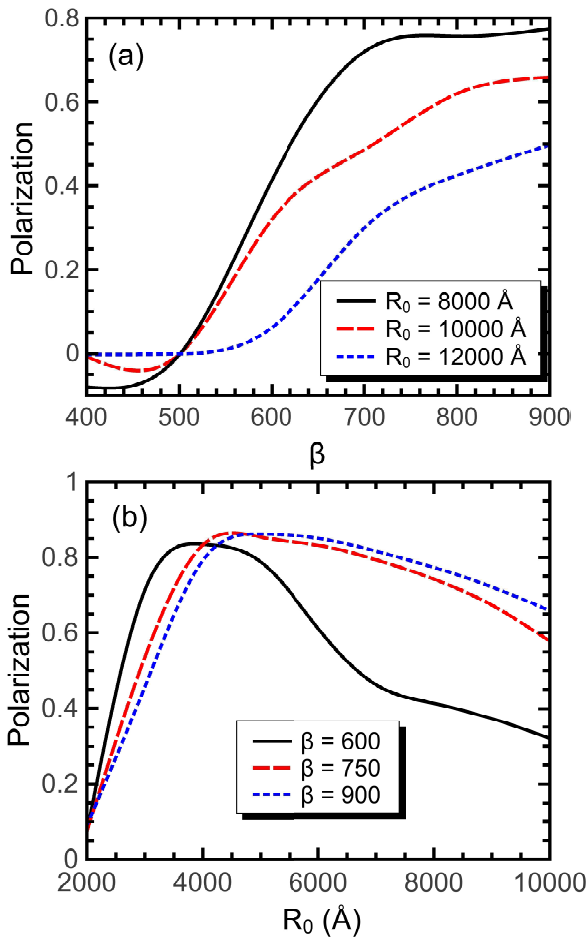}}
\caption{(Color online) Valley polarization of the outgoing wavepacket, with $k = 0.06$ \AA$^{-1}$, as a function (a) of the width of the strained region $\beta$, for different radii $R_0$, and (b) as function of the strain radius $R_0$, for different $\beta$ values.} \label{fig:polarization}
\end{figure}

Such a picture of snakes states strongly suggest a valley filtering effect. In fact, if one now considers a system with finite strain region $\beta = 900$ (in units of the inter-atomic distance $a_0 = 1.42$ \AA) and $R_0 = 10,000$ \AA, a wavepacket with $k = 0.06$ \AA$^{-1}$ around the $K$ cone passes through this region with a high probability $P_3 \approx 0.9$, whereas the same packet in $K'$ would have a much lower transmission probability $P'_3 \approx 0.3$. 

\begin{figure}[!b]
\centerline{\includegraphics[width=0.9\linewidth]{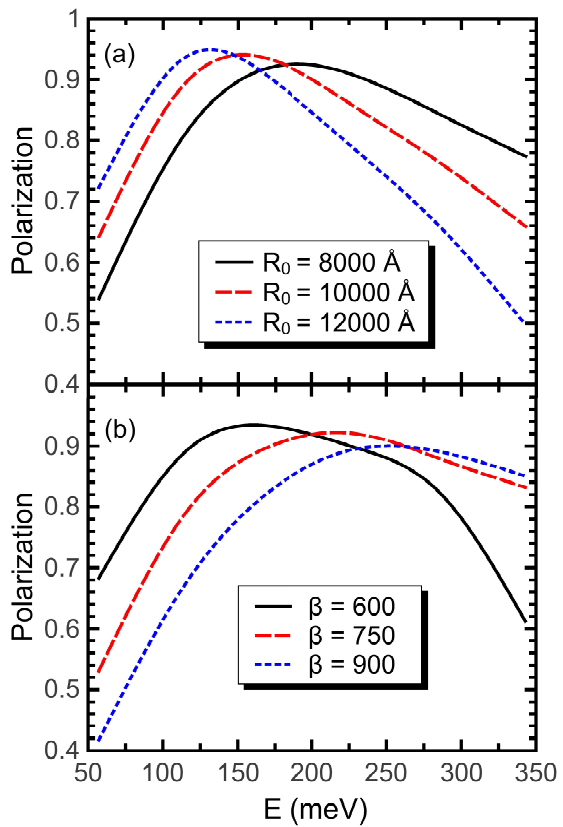}}
\caption{(Color online) Valley polarization of the outgoing wavepacket as a function of its energy, considering (a) $\beta = 900$ and different radii $R_0$, and (b) $R_0 = 6,000$ \AA, for different values of $\beta$.} \label{fig:polarizationE}
\end{figure}

Let us now search for an optimization of the valley polarization effect. The polarization, as defined by $P = 1-P_3/P_3^{\prime}$, where $P_3^{\prime}$ is the transmission probability for a wavepacket in the $K'$ cone, is shown in Fig. \ref{fig:polarization} for $k = 0.06$ \AA$^{-1}$, assuming different parameters $\beta$ and $R_0$. For a fixed strain radius $R_0$, increasing the length of the strain region $\beta$ increases the polarization, as shown in Fig. \ref{fig:polarization}(a). Besides, results in this panel also suggest that decreasing $R_0$ would always improve the polarization; this would be reasonable, since smaller radii yield stronger distortions in the lattice and, consequently, larger pseudo-magnetic fields. This is however not always the case: Fig. \ref{fig:polarization}(b) shows that even for $\beta$ as large as $900$ \AA, decreasing the strain radius will always lead to a maximum polarization at an intermediate value $R_0 \approx$ $5,000$ \AA, so that the polarization is reduced as the radius is further decreased. This is due to the fact that the smooth connection between the unstrained ribbon leads and the strained region might end up creating a complicated pseudo-magnetic field distribution, with regions with fields pointing to opposite directions, which would harm the polarization effect investigated here. This also explains the negative polarization observed for small $\beta$ in Fig. \ref{fig:polarization}(a). 

So far, all results were obtained for $k = 0.06$ \AA$^{-1}$, which corresponds to a wavepacket energy $E = 343$ meV. It is however important to check how the polarization depends on the wavepacket energy. This is shown in Fig. \ref{fig:polarizationE}, where we verify that the valley filtering process proposed here has an optimal range of energies. Indeed, if the energy is too low, the pseudo-magnetic Lorentz orbits would have a very small radius, so that only portions of the wavepacket that are very close to the $y = 0$ line would pass through the system as snake states, whereas the rest of the wavepacket readily turns back. On the other hand, if the energy is too high, orbit radii may end up being larger than the strained region length, so that the snake-like propagation that leads to valley polarization no longer occurs. Moreover, the Lorentz orbit radius is inversely proportional to the pseudo-magnetic field intensity, therefore, increasing the strain by reducing $R_0$ would also lead to Lorentz orbits with smaller radius. This classical picture is consistent with our numerical findings: in Fig. \ref{fig:polarizationE}(a), the largest strain radius $R_0 = 12,000$ \AA\ provides the fastest decay of polarization as the energy increases, since it yields a lower strain-induced pseudo-magnetic field and, therefore, energies slightly higher than the optimal $E \approx 125$ meV already provide orbit radii larger than the strained region length. Conversely, for energies lower than $E \approx 125$ meV, $R_0 = 12,000$ \AA\ provides the best polarization, as its weaker pseudo-magnetic field compensates for the low energy and prevents the orbits radius of becoming too small. Also, Fig. \ref{fig:polarizationE}(b) shows that, for higher energies, where orbit radii are larger, polarization is more efficient for larger length $\beta$, and the energy for optimal polarization increases with this parameter.

The valley filtering effect by pseudo-magnetic kinks demonstrated here is also expected to be robust against impurity and defects scattering: as already discussed, electrons in each valley have only one possible direction of propagation (see Fig. \ref{fig:snakes}), due to the monotonic behavior of all the energy states as a function of momentum (see Fig. \ref{fig:spectrum}), which provides a single direction for the group velocity in each valley. After scattering by impurities or defects, the electron must end up in one (or a combination) of the states in Fig. \ref{fig:spectrum}. If the electron is already in the valley that allows its propagation through the system (as a snake state), with positive velocity (i.e. monotonically increasing energies as a function of momentum), any component of the scattered electron wave function that ends up in the other valley must be deflected backwards, since there is simply no energy state in that valley with positive velocity as well. Thus, provided the pseudo-magnetic field kink distribution is preserved, only electrons in one of the valleys are allowed to reach the other end of the ribbon, even after scattering events.

\section{Conclusions}

We have investigated the wavepacket propagation through a graphene nanoribbon with armchair edges for a specific strain distribution. The latter provides a pseudo-magnetic barrier kink along the ribbon. By following the trajectory of the center-of-mass of the wavepacket, calculated by solving the time-dependent Schr\"odinger equation for the tight-binding Hamiltonian, one observes snake states, which have a fixed propagation direction, consistent with the pseudo-magnetic kink picture. However, one can analytically verify that, by reversing the magnetic kink, the propagation direction of snake states must be reversed. 

Since the pseudo-magnetic field points towards opposite direction in the different Dirac cones, wavepackets in the different cones can only have fixed opposite directions of propagation. This effect results in an efficient valley filtering process, which does not require either lattice defects, edge engineering, or any externally applied fields or potentials. Our numerical results show significant valley polarization through this system, which can be optimized by the parameters ($\beta$, $R_0$) that depend on the electron energy (i.e. the Fermi energy). 

Notice that the in-plane circular deformation of a graphene nanoribon proposed here is just one particular way of inducing a kink pseudo-magnetic field barrier: any other strain distribution that produces such a pseudo-magnetic kink would lead to similar valley filtering effect, which requires only a graphene ribbon with armchair boundaries (to avoid edge propagation) and a pseudo-magnetic field that flips its direction across a line parallel to them.

\acknowledgments
Discussions with R. Grassi are gratefully acknowledged. This work was supported by the Brazilian Council for Research (CNPq), under the PRONEX/FUNCAP and Science Without Borders (SWB) programs, CAPES, the Lemann Foundation, and the Flemish Science Foundation (FWO-Vl).

\end{document}